\documentclass[twocolumn,showpacs,preprintnumbers,amsmath,amssymb]{revtex4}

\usepackage{graphicx}
\usepackage{dcolumn}
\usepackage{bm}

\begin{document}


\title{Forbidden patterns in financial time series}

\author{Massimiliano Zanin}
\email{massimiliano.zanin@hotmail.com}
\affiliation{%
Universidad Aut\'onoma de Madrid, 28049 Madrid, Spain
}%

\date{\today}

\begin{abstract}
The existence of {\em forbidden patterns}, i.e., certain missing sequences in a given time series, is a recently proposed instrument of potential application in the study of time series. Forbidden patterns are related to the permutation entropy, which has the basic properties of classic chaos indicators, such as Lyapunov exponent or Kolmogorov entropy, thus allowing to separate deterministic (usually chaotic) from random series; however, it requires less values of the series to be calculated, and it is suitable for using with small datasets. In this Letter, the appearance of forbidden patterns is studied in different economical indicators like stock indices 
(Dow Jones Industrial Average and Nasdaq Composite), 
NYSE stocks (IBM and Boeing) and others (10-year Bond interest rate), to find evidences of deterministic behavior in their evolutions. Moreover, the rate of appearance of the forbidden patterns is calculated, and some considerations about the underlying dynamics are suggested.
\end{abstract}

\pacs{89.65.Gh,05.45.Tp}

\maketitle


Extracting information from real time series has been a hot topic during the
last decades \cite{box70,got81,bro87,mon90,eat91,sch07,ham94}. From the point of view of time series analysis
two main goals arise when facing a real evolution of a certain variable:
first, identifying the underlying nature of the phenomenon represented by the sequence of 
observations and, second, trying to predict the evolution of the variable. Both of these
goals, identification and forecasting, require the treatment of the time series, usually by combining 
different tools. Statistical methods in order to obtain a model
of the mean process have been the classical approach, leading to autoregressive, integrated and moving
average models \cite{bro87}. Characterization of non-linear time series, mainly chaotic, has also
attracted the interest of the scientific community \cite{aba93,kan04}, where phase space reconstruction,
spectral analysis or wavelets methods have been revealed to be good indicators of the 
underlying dynamics of a real time series.

Recently the study of the {\em order patterns} has been proposed as a technique of
evaluating the determinism of a given time series \cite{ami05,ami06,ami07}. 
Consider a discrete information source emitting a series of observable values $[x_1, x_2, \ldots, x_N]$, ordered by time;
it is possible to split the data in overlapping sets of length $d$,
and study their order patterns, e.g., $x_1<x_2<...<x_d$. 
Every group of $d$ adjacent values form a certain permutation $\Pi$, which is one of the $d!$ possible permutations \cite{ami07}. 
The basis of the 
topological permutation entropy \cite{ban02} is to define, for every group of $d$ adjacent values within a discrete dataset, the corresponding 
permutation pattern, and 
study the overall statistics of these patterns \cite{ami05,ami06}.
For a pattern dimension of, e.g., $d = 3$, if $x_2 < x_1 < x_3$, the resulting pattern would be $\Pi = (2, 1, 3)$, i.e., the lower 
value of the series is the second, followed by the first, while the last value is the highest. 
In principle, the dimension $d$ could be any integer higher 
than one and several values of $d$ could be defined within the same dataset. 
We just have to take into account that the higher the dimension considered, the larger the quantity of data needed, 
so in this paper we consider $3 < d < 7$, which are high enough values to distinguish the existence (or not) of forbidden patterns.

When analyzing a time series of length $N$, we obtain $N-d+1$ overlapping groups of adjacent values, 
each one with a corresponding order pattern. If the series has a random behavior,
any permutation can appear, and therefore no pattern is {\it forbidden}. Moreover, 
their probability distribution should be flat since any permutation has the same probability of occurrence when 
the dataset is long enough to exclude statistical fluctuations.
Nevertheless, when the series corresponds to a chaotic variable
there are some patterns that cannot be encountered in the data
due to the underlying deterministic structure:
they are the so-called {\it forbidden patterns}. 
It has been demonstrated that most chaotic systems exhibit forbidden patterns, and that in 
many cases (ergodic finite-alphabet information sources) 
the measure of the number of this patterns 
is related to other classic metric entropy rates (e.g., the Lyapunov exponent) \cite{ami07}.

In the current Letter we study the existence of forbidden patterns in economical time series. Specifically,
we analyzed the appearance (or absence) of this patterns in several financial indicators. We have
seen that, despite quantitative differences, the order pattern analysis reveals the existence of forbidden order patterns in all
time series analyzed here, which indicates an underlying deterministic behavior. Furthermore, we have followed the evolution 
of the forbidden patterns, which allows to identify periods of time where noise or randomness is overtaking 
the deterministic behavior of the financial indicators.

\begin{figure}[!htb]
\includegraphics[width=8.3cm]{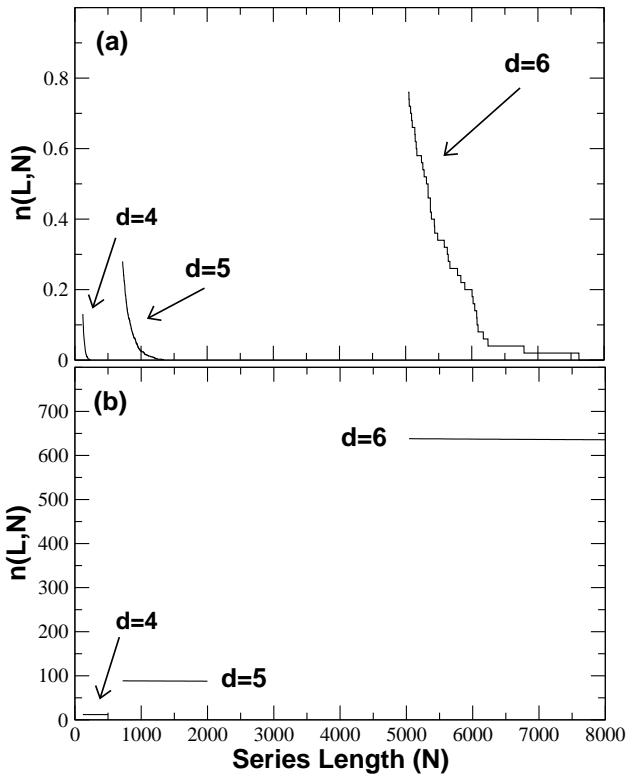}
\caption[sdh]{
Number of {\it forbidden patterns} for (a) a random series and (b) a deterministic series 
generated by the logistic map. Note how the deterministic data generates a higher number of forbidden patterns, and how its number does not depend on the series length.} 
\label{fig:f01}
\end{figure}

In order to illustrate the absence of certain order patterns in deterministic time series, we plot in Fig.\ref{fig:f01} the number of existing forbidden patterns in a random
time series (a) and an equivalent series generated by a logistic map (b), which is obtained from $x_{n+1}=4x_n(1-x_n)$ and  
$0 \leq x_0 \leq 1$.
At first sight, we can see how the logistic function generates a higher number of forbidden patterns, 
revealing its chaotic behavior. Moreover, the existence of this kind of patterns is intrinsic to its nature (i.e., its construction): adding samples to the series does not help in lowering their number. On the other hand, the random case shows forbidden patterns due only to statistical limitations, and their number goes to zero for long enough time series. 
From the example above, we can construct a rule for time series characterization: if the number of forbidden patterns 
is greater than the quantity encountered in a equivalent random series 
(at least an order of magnitude above), that series has a deterministic structure \cite{ami07}. 

As we see in Fig. \ref{fig:f01}-(a) it is important to define long enough series to avoid statistical distortions at the output. 
For a pattern dimension $d$ and a time series of length $N$; the number of possible patterns is $d!$, while the number of groups of data of dimension $d$ is $N-d+1$. To guarantee that every pattern appears at least once in the dataset, we must ensure that $N-d+1 > d!$, and therefore that $N > d! + d - 1$. For that reason, we have chosen a lower bound of $N>(d+1)!$, which avoids problems of undersampling.

\begin{figure}[!bt]
\includegraphics[width=8cm]{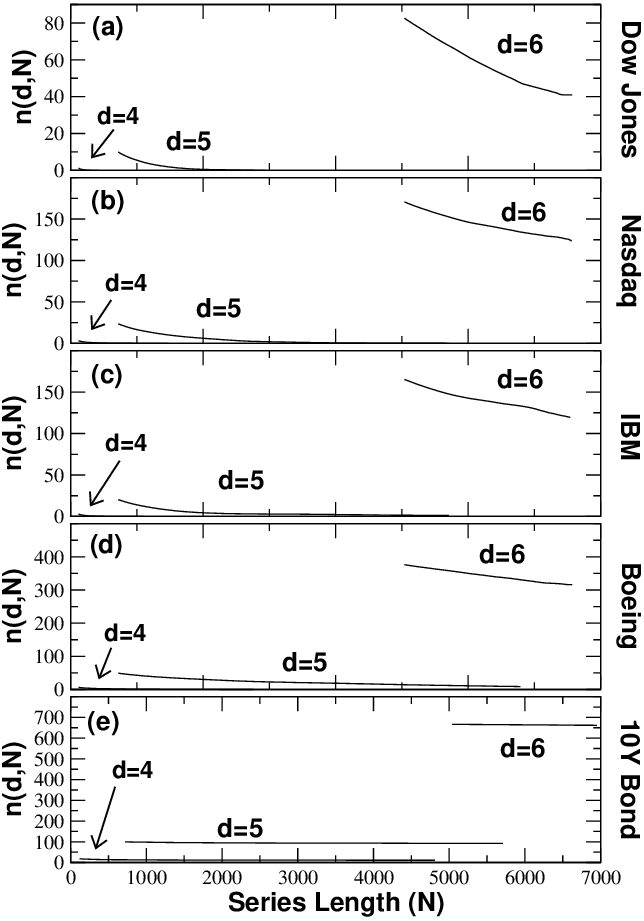}
\caption[sdh]{
Number of forbidden patterns $n(d,N)$ for: Dow Jones Industrial Average (a), Nasdaq Composite (b), NYSE IBM (c), NYSE Boeing (d), 10-Years Bond interest rate (e). Three different values of the pattern length $d$ are computed, $d=\{4,5,6\}$.} 
\label{fig:f02}
\end{figure}

We have studied the appearance of forbidden patterns in real the time series of different financial indicators, namely: Dow Jones Industrial Average \cite{dj}
and Nasdaq Composite \cite{nas} (US stocks indices), IBM and Boeing \cite{nys} (NYSE stocks), and the 
10-Year U.S. Bond rate \cite{10y}.
As in previous examples we have computed the $d=\{4,5,6\}$ ordinal patterns for different lengths of the time series.
Results are shown in Figs. \ref{fig:f02}-(a)-(b)-(c)-(d)-(e), while Fig. \ref{fig:f01}-(a) corresponds to an equivalent time series
that has been randomly generated.

At first sight it is clear that in all cases the number of forbidden patterns are much higher (two orders of magnitude) than 
the expected in the random case. This fact reveals the existence of a deterministic component in the evolution of all
time series which coexists with stochastic fluctuations. 
As shown in \cite{ami07}, the combination of a deterministic and noisy signal introduces a decay in the number
of forbidden patterns when increasing the length of the dataset, which is the case of the financial time series. 
Nevertheless, it has been demonstrated that the number of forbidden patterns is at least
one order of magnitude higher than in the random case, even for moderate values of the noisy signal \cite{ami07}. In this sense, we obtain
in all cases a number of forbidden patterns that is at least two orders of magnitude from that of the random case, indicating
the existence of driving deterministic forces.
From (a) to (e) we have ordered the financial indices according to their number of forbidden patterns, e.g., from the least deterministic
to the most deterministic series. In this way, we can infer what time series are more influenced by random fluctuations, which
is a valuable information for the development of economic models that predict the evolution of the financial time series.
Moreover, the presence of forbidden patterns seems to be related with the market operations involving the first four series: their mean trade volume is ordered from high to low (respectively 2.5 B\$, 2 B\$, 5 M\$ and 4 M\$), as if a great number of operations results in a more random behavior.
It is worth mentioning that the decay in the number of forbidden patterns with the series length decreases with the total number
of forbidden patterns, since it is somehow related: the lower the dependence on the time series length, the higher the deterministic
part of the series. 
Finally we must remark the results obtained for the 10-year Bond rate. The high number of forbidden patterns and the fact
that they are nearly independent from the length of the time series indicates the high deterministic behavior of this
particular financial series. 


At this point, let us move our attention to the probability distribution function of the order patterns
within the time series.
When dealing with random time series, every permutation pattern should have the same probability to appear, and
therefore, when $N \rightarrow \infty$ their probability 
distribution should be a flat function. In limited (random) time series
the probability should follow a Poisson distribution, centered at $n_{mean}=T_p/N_p$, where $N_p$ is the total number of 
sequences and
$T_p$ is the number of possible patterns.
On the other side, we have seen that deterministic time series have certain forbidden patterns, which
may lead to different probability distributions. Particularly, in cases where noise is mixed with a deterministic signal, 
the analysis of the probability distribution can show interesting results, specially if it moves away from the poissonian distribution.

\begin{figure}[!bt]
\includegraphics[width=8cm]{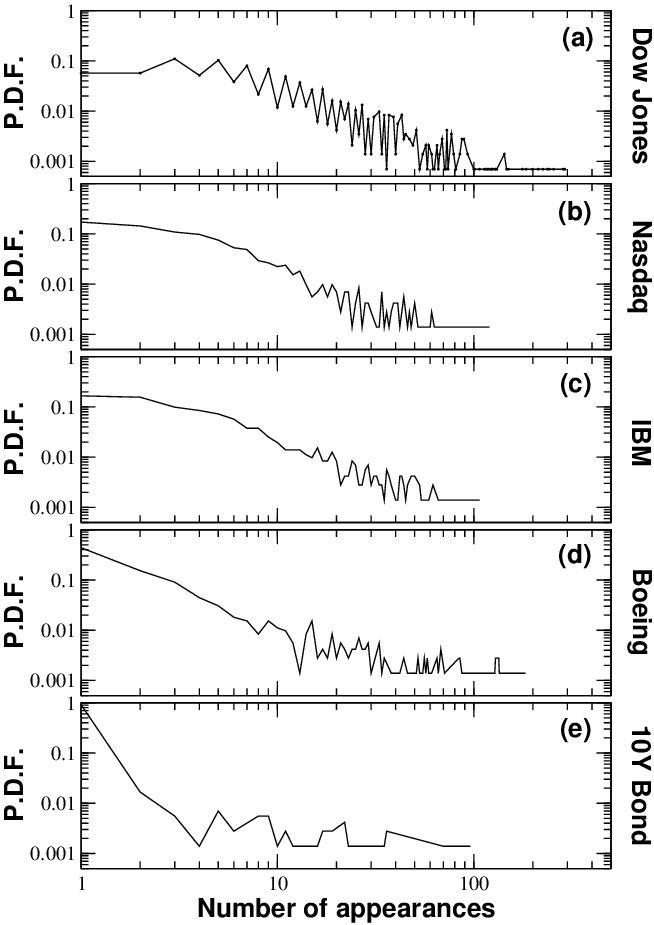}
\caption[sdh]{
Probability distribution function (P.D.F.) of the order pattern occurrence, i.e., the probability
of finding a pattern repeated certain number of times, for the financial series of Fig. \ref{fig:f02}. Note the logarithmic
scale at both axis, which reflects the heterogeneity of the probability.} 
\label{fig:f03}
\end{figure}

\begin{figure*}[!tb]
\centerline{
\includegraphics[width=16cm]{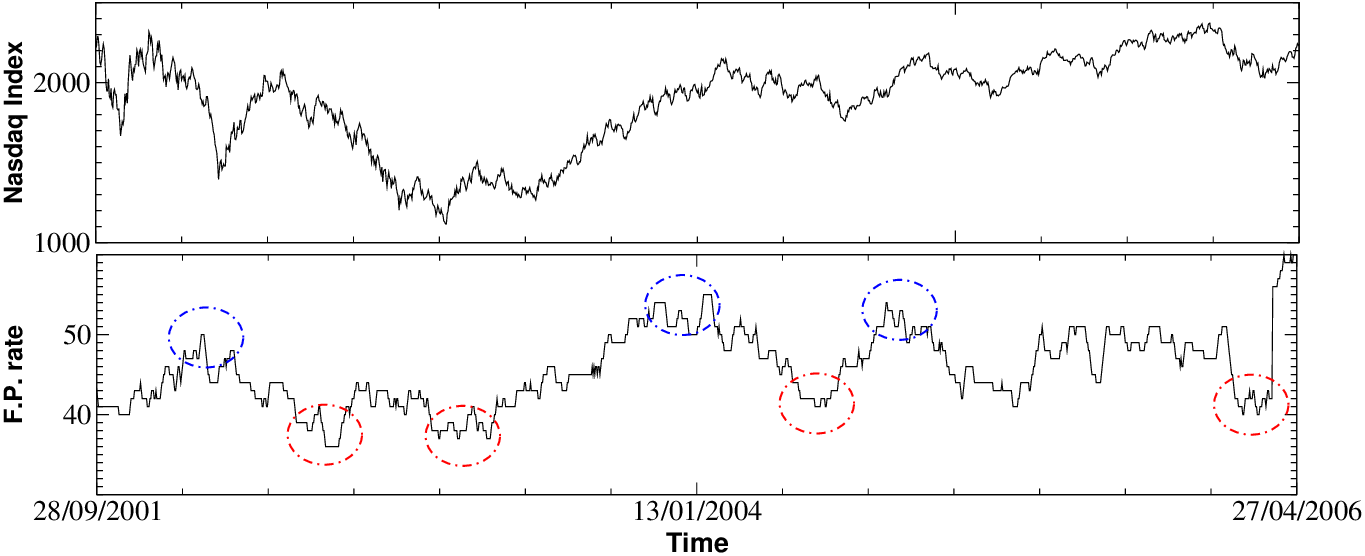}
}
\caption[sdh]{
Evolution of the Nasdaq Index and the corresponding forbidden patterns (F.P.) of order 5 ($d=5$), for a time
window of 200 samples. 
Blue and red ellipses indicate the existence of relative maxima/minima, which is related, respectively, with
the increase and decrease of the predictability of the system.
} 
\label{fig:f04}
\end{figure*}


Figure \ref{fig:f03} shows the results for the financial series studied here. 
We can see how, in all series, the probability distribution show
heavy tails, which indicates the heterogeneity in the probability of appearance (i.e., non-poissonian distribution). 
This fact distinguishes again the financial time
series from the random ones, and shows that the study of the probability distribution function of the order patterns could be
a good indicator of the deterministic nature of the series. 
Since series of Fig. \ref{fig:f03} are ordered from low to high deterministic behavior, it is interesting to note
that the number of patterns that appear only once or twice increases with the determinism of the
series.



Finally, we will have a look at the time evolution of the forbidden patterns in order to see if their number shifts in time.
Fig. \ref{fig:f04} represents the evolution of the Nasdaq index from September, 28th 2001 to April, 27th 2006 (upper plot), 
together with the evolution of their forbidden patterns (bottom plot). 
In the latter, each point of the plot represents the number of forbidden patterns of dimension $d=5$ for a time window of 200 samples. 
We can observe how the number of forbidden patterns do not have a constant rate, on the contrary, it fluctuates in time. Its
number is a good indicator of the randomness of the system at a precise moment. In this way, a decrease of the rate of 
forbidden patterns reflects that random forces are increasing (red ellipses in Fig. \ref{fig:f04}), leading to an enhance of the unpredictability of the system. On the contrary,
when forbidden patterns increase (blue ellipses in Fig. \ref{fig:f04}), randomness is decreasing, which would indicate that 
the evolution of the time series
would be more predictable. Note that we are not obtaining information about how the system is going to evolve, since we do not
have a model describing the time series, but if we had it, the time periods where the number of forbidden patterns increase, would
be the most suitable to apply it. Similar results, not shown here, are obtained for the rest of the financial time series.




In summary, we have studied the existence of forbidden order patterns in financial time series. We have seen
that a high number of forbidden patterns, two orders of magnitude higher than in random series, reveals an underlying deterministic
behavior in the series, which indicates that predicting models could be suitable to forecast the behavior
of financial time series. Furthermore, we have shown that the forbidden pattern rate could be an appropriate
tool to quantify the randomness of certain time periods within the financial series, and therefore
evaluate the uncertainty of the market. To our knowledge, these are the first results of the forbidden pattern
evaluation in real time series, which also open interesting questions to be addressed in the future. 
We believe that the application of this technique to real time series of different nature, e.g., biological or medical datasets, 
will show promising results. 


The author is indebted to Javier M. Buld\'u for his helpful comments and suggestions.



\end{document}